\documentclass[12pt]{iopart}

\usepackage{graphicx}
\begin{document}

\title{Universality of transport properties of ultra-thin oxide
films}

\author{V Lacquaniti$^1$, M Belogolovskii$^2$, C Cassiago$^1$, N De Leo$^1$, M Fretto$^1$, A Sosso$^1$}

\address{$^1$ National Institute of Metrological Research, Electromagnetism Division,
Strada delle Cacce 91, 10135 Torino, Italy}

\address{$^2$ Donetsk Institute for Physics and Engineering, National Academy of
Sciences of Ukraine, Str. Rosa Luxemburg 72, 83114 Donetsk,
Ukraine}

\ead{v.lacquaniti@inrim.it}

\begin{abstract}
We report low-temperature measurements of current-voltage
characteristics for highly conductive Nb/Al-AlO$_x$-Nb junctions
with thicknesses of the Al interlayer ranging from 40 to 150 nm
and ultra-thin barriers formed by diffusive oxidation of the Al
surface. In a superconducting state these devices have revealed a
strong subgap current leakage. Analyzing Cooper-pair and
quasiparticle currents across the devices, we conclude that the
strong suppression of the subgap resistance comparing with
conventional tunnel junctions is not related to technologically
derived pinholes in the barrier but rather has more fundamental
grounds. We argue that it originates from a universal bimodal
distribution of transparencies across the Al-oxide barrier
proposed earlier by Schep and Bauer. We suggest a simple physical
explanation of its source in the nanometer-thick oxide films
relating it to strong local barrier-height fluctuations in the
nearest to conducting electrodes layers of the insulator which are
generated by oxygen vacancies in thin aluminum oxide tunnel
barriers formed by thermal oxidation.
\end{abstract}

\submitto{\NJP}
\maketitle

\section{Introduction}
During the past decade, there has been considerable progress in
understanding fundamental electronic properties of ultra-thin
dielectrics, in particular, charge transport across disordered
oxide layers. The increasing interest in such materials has been
motivated by their promising applications as a gate dielectric in
metal-oxide-semiconductor transistors instead of SiO$_2$ and a
blocking dielectric for new-generation flash memory cells (see the
papers~\cite{Perevalov,Arhammar} and references therein) as well
as a potential barrier for current carriers in superconducting
tunnel junctions~\cite{Blamire,Greibe}. In particular, it relates
amorphous nanometer-thick alumina AlO$_x$. Unfortunately, in
practice, high demands on the ultra-thin oxide films as
current-blocking high-$k$ dielectrics with a minimal dissipation
are not usually fulfilled and a strong leakage current through the
amorphous oxides is emerging as a potential limitation for most
novel applications ~\cite {Perevalov,Arhammar,Greibe}. This
finding has been often attributed to microscopic defects in the
tunnel barrier with a greatly enhanced local
transparency~\cite{Schrieffer}, commonly known as "pinholes". It
has been assumed that the pinholes shunt the tunnel current and
thus dominate the charge transport across the junctions. But a
recent extensive analysis of a differential subgap conductance in
highly resistive superconducting tunnel junctions with Al-oxide
barriers~\cite{Greibe} has ruled out the pinholes as the origin of
the excess current, at least, in samples with normal conductance
$G_N$ per geometric area $A$ between 10$^4$ and
10$^7$~(Ohm$\cdot$cm$^2$)$^{-1}$. Hence, the identification of
charge transport mechanism in ultra-thin oxide films as well as
the defects nature in amorphous dielectrics like AlO$_x$ remains
to be of great scientific and practical interest.

In this paper, we study charge transport across extremely thin
AlO$_x$ layers sandwiched between two metallic electrodes with
specific normal conductance ranged in the interval from
2$\times$10$^7$ to 7$\times$10$^7$~(Ohm$\cdot$cm$^2$)$^{-1}$. It
means that in our case the barrier thicknesses were significantly
reduced comparing to those in the paper~\cite{Greibe} and, hence,
the presence of pinholes was more probable. Below, we not only
reject the hypothesis about dominating role of microscopic
pinholes in the ultra-thin oxide layers but use our experimental
findings for four-layered Nb/Al-AlO$_x$-Nb junctions to show that
the subgap resistance suppression in a superconducting state
comparing to conventional tunnel junctions originates from a
universal bimodal distribution of transparencies across the oxide
barrier.

Concerning the latter statement, we should note that in the
general case quantum transport through a mesoscopic physical
system is determined by different factors like the system
dimensionality, its size and shape, carrier density, and other
sample-specific features. Universal transport properties, if they
are, should be independent from microscopic details of particular
materials and may include only a limited number of macroscopic
characteristics. Inter alia, the universality can be expected for
"dirty" systems with a very large spread of random potentials
scattering current carriers. In the seminal paper by Schep and
Bauer~\cite{Schep} the authors studied quantum transport
properties of a single perfectly disordered interface (I) whose
conductance is much smaller than conductance of a ballistic
conductor of the same cross-section and whose thickness $d$ is
sufficiently shorter than the Fermi wavelength $\lambda _{\rm F}$
in metallic bulks around it. They found that the distribution
$\rho (D)$ of eigenvalues $D$ of a transfer matrix connecting
incoming and outgoing electronic modes (it is the product of the
related transmission amplitude matrix and its Hermitian conjugate)
does belong to a universality class and is given by an expression
\begin{equation}
\rho (D)=\frac{\hbar G_{\rm N}}{e^2}\frac{1}{D^{3/2}(1-D)^{1/2}}.
\end{equation}
with $G_{\rm N}$, the disorder-averaged conductance. For a sample
with such an interface any physical quantity $f$ described by a
linear statistics $f(D)$ (like conductance or shot noise power)
can be calculated as $\int\limits_0^1 {f(D)\rho (D)dD}$.

The best way to study the distribution $\rho (D)$ in a
nanometer-thick insulating film is to place it between two
metallic layers and to measure current-voltage $I-V$
characteristics of the tunnel junction at very low temperatures
when one or both electrodes are transformed into a superconducting
(S) state. The reason consists in the fact that the shape of
quasiparticle $I-V$ curves for SIS trilayers with a single quantum
channel is extremely sensitive to the transmission
probability~\cite{Wolf} and, if the total current across the
junction can be represented as a sum of independent contributions
from individual transverse modes, its voltage dependence would be
definitely determined by the function $\rho (D)$.

A first analysis of the $\rho (D)$ distribution in
subnanometer-thick AlO$_x$ layers using a superconducting
heterostructure was performed by Naveh {\it{et al.}}~\cite{Naveh}
who measured current-vs-voltage and differential
conductance-vs-voltage characteristics of planar highly conductive
Nb-AlO$_x$-Nb trilayers at 1.8 K. In spite of the presence of an
AlO$_x$ barrier, the quasiparticle $I_{\rm {qp}}-V$ curves did not
exhibit typical for conventional tunnel junctions
behavior~\cite{Wolf} with $R_{{\rm{sg}}} \gg R_{\rm{N}}$. Here
$R_{\rm{sg}}$ is the junction differential resistance at voltages
$\left| V \right| < 2\Delta /e$, known as a subgap resistance,
$\Delta $ is the energy gap of a superconducting electrode, in
this case, niobium, $R_{\rm{N}}$ is the trilayer normal-state
resistance. On the contrary, the ratio $R_{\rm{sg}} /R_{\rm{N}}$
in the state when the Josephson critical current was totally
suppressed by magnetic fields was near unity. Moreover,
conductance-voltage curves exhibited well pronounced subharmonic
gap structure known as a fingerprint of multiple Andreev
reflections in high-transparent SIS
samples~\cite{Bratus,Averin,Bardas}. All these findings were
explained~\cite{Naveh} as an impact of a broad double-peak
distribution of Al-oxide barrier transparencies described by Eq.
(1).

It should be noticed that the presence of a striking,
polarity-independent deviation from standard single-particle
tunneling $I-V$ curves in SIS junctions, is well known from 60s of
the last century~\cite{Taylor}. The subgap leakage was observed
not only in highly conductive junctions~\cite{Blamire} but also in
low-transmission Nb-AlO$_x$-Nb trilayers~\cite{Milliken} and in
both cases evidences of an important role played by multiple
Andreev-reflection processes have been
reported~\cite{Naveh,Greibe} although up to now it is not clear if
their nature is identical or not~\cite{Bezuglyi}. Recently, there
has been a resurgence of interest in this effect since enhanced
quasiparticle subgap conductance within individual junctions can
cause decoherence effects in Josephson-effect-based
superconducting quantum devices~\cite{Im}. The "quality factor" of
an SIS junction, which is usually defined by the ratio
$R_{\rm{sg}}/R_{\rm{N}}$, is also important for low noise
performance of SIS mixers and detectors where it is required to be
at least $\sim15$ for low noise performance~\cite{Endo}. At the
same time, to achieve a broad rf-bandwidth, the junction must have
a very high critical-current density $J_{\rm c}$ or a small
$R_{\rm N}A$ product where $A$ is the junction
area~\cite{Kawamura}. The two requirements contradict each other
because of a strong positive correlation between the "junction
quality" and its specific conductance~\cite{Blamire}. Another
aspect of the same problem relates to the implementation of
Josephson junctions operating in an overdamped mode when the ratio
$R_{\rm{sg}}/R_{\rm{N}}$ should be suppressed as much as
possible~\cite{LacquanitiJAP}. Since thermally oxidized aluminum
oxide tunnel barriers provide the most reliable junction
fabrication technology due to its reasonably good properties and
comparatively easy fabrication~\cite{Blamire}, more fundamental
knowledge about possibility to change the ratio
$R_{\rm{sg}}/R_{\rm{N}}$ in a controllable way is needed in order
to provide more insight into intrinsic structure of the AlO$_x$
insulating layer~\cite{Tan} as well as into the nature of current
fluctuations in related superconducting
devices~\cite{Julin,Faoro}.

Earlier investigations of subgap processes have been mainly
concentrated on symmetric S-I-S~\cite{Blamire} and asymmetric
S-I-N~\cite{Zorin} junctions. At the same time, it was
stressed~\cite{Ternes} that experimental studies of asymmetric
S$_1$-I-S$_2$ devices can provide more information since the
structure of I-V curves at voltages below $\left ( \Delta _1 +
\Delta _2  \right)/e$ is more rich and
pronounced~\cite{Hurd1996,Hurd1997}. The aim of this work is to
reexamine the conclusion of the paper~\cite{Naveh} about very
large variations of local barrier transparencies in
nanometer-thick AlO$_x$ layers, using asymmetric Josephson
junctions with a proximity-coupled S/N bilayer as one of the
device electrodes and studying the thermal effect in SNIS devices.
Comparing with our previous
experiments~\cite{Lacquaniti2007,LacquanitiJPCS,LacquanitiJAP}, we
have enlarged the temperature interval studied, performing precise
measurements of dissipative transport characteristics at
temperatures significantly below and above 4.2 K. We have
succeeded to extract corresponding $R_{\rm{sg}}/R_{\rm{N}}$ values
and to compare them with our numerical simulations which are based
on Eq. (1) applied to a more complicated than before four-layered
junction geometry. Besides it, we present the data of novel
rf-measurements proving the Josephson character of charge
transport across our samples in a superconducting state. The rest
of the paper is devoted to a simple physical interpretation of the
Schep-Bauer distribution (1), the primary focus of the paper,
relating it to strong local barrier-height fluctuations generated
by oxygen vacancies in extremely thin Al-oxide films.

\section{Experimental}
We present experimental results obtained on asymmetric
Nb/Al-AlO$_x$-Nb junctions developed at Instituto Nazionale di
Ricerca Metrologica~\cite{Lacquaniti2007}. The main differences
with a standard Nb-technology~\cite{Gurvitch} are the following:
the thickness of the Al interlayer $d_{\rm {Al}}$ was increased
from 5-10 nm up to 40-150 nm whereas the exposure dose, the
product of the oxygen pressure and the oxidation time, was
decreased from more than 1000 Pa$\cdot $s down to 150-500 Pa$\cdot
$s (note that in the paper~\cite {Greibe} oxygen doses varying
from 1.7$\times$10$^3$ to 8$\times$10$^5$ Pa$\cdot $s produced
extremely thick barriers). Because of comparatively small exposure
doses in our case, specific normal conductances of the
Nb/Al-AlO$_x$-Nb devices were as high as (2 -
7)$\times$10$^7$~(Ohm$\cdot$cm$^2$)$^{-1}$ which is of the same
order of magnitude as those of highly-conductive symmetric SIS
Josephson junctions studied by Naveh {\it{et al.}}~\cite{Naveh}.
Other details of our 25 $\mu$m$^2$-area junctions are given
elsewhere~\cite{LacquanitiJAP}.
\begin{figure}[ht]
\begin{center}
\includegraphics[width=300pt]{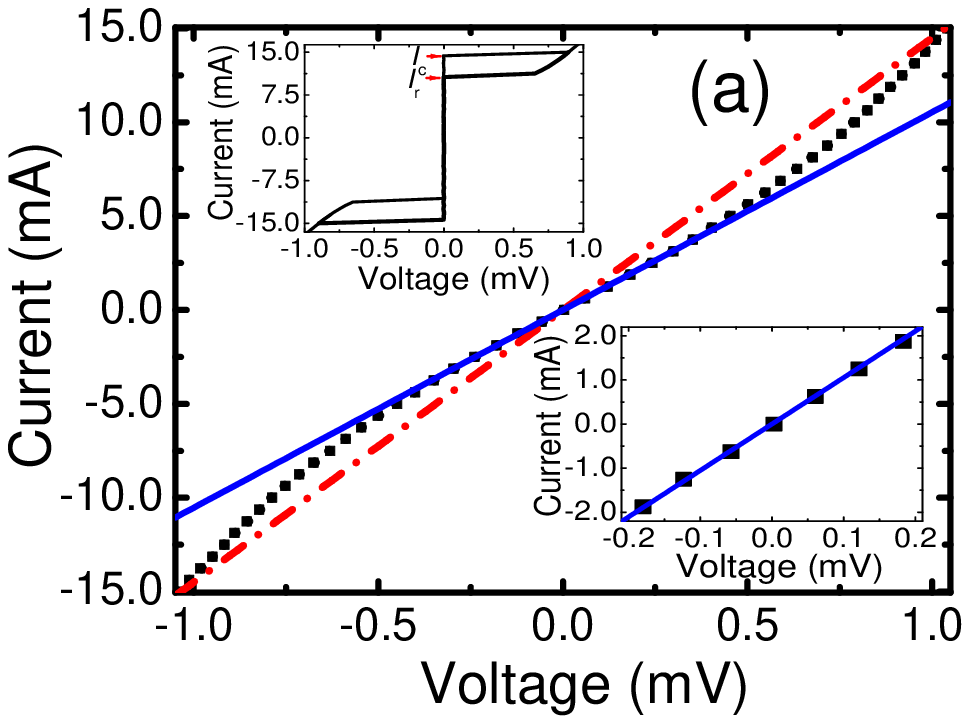}\\
    \includegraphics[width=300pt]{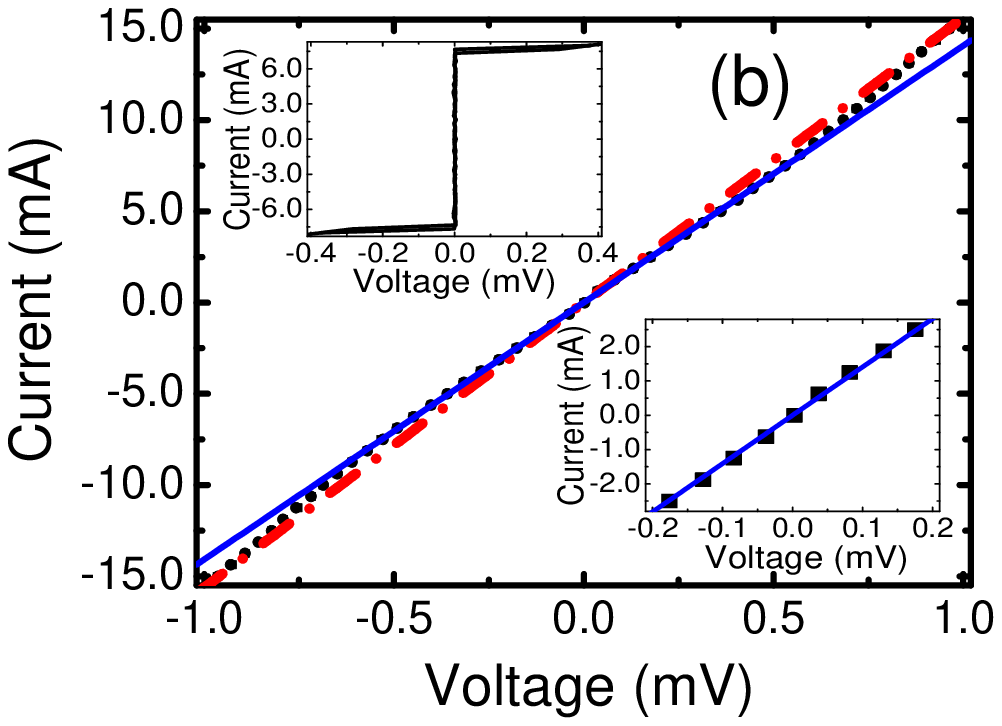}\\
\caption { Dissipative current-voltage characteristics of a
representative Nb/Al-AlO$_x$-Nb junction with $d_{\rm {Al}}=110$
nm at 1.7 K (a) and 4.3 K (b) measured in external magnetic fields
about 50 mT applied to the tunnel sample through a suitable coil.
Left insets demonstrate hysteretic behavior of the supeconducting
current with switching $I_{\rm {c}}$ and retrapping $I_{\rm {r}}$
values whereas the main panels show quasiparticle $I_{\rm {qp}}-V$
characteristics with suppressed Cooper-pair contribution (dotted
curves). Right insets exhibit the same curves in an enlarged scale
near $V=0$. Solid and dashed-dotted straight lines correspond to
Ohm's laws with $R_{\rm{sg}}=95$ mOhm and $R_{\rm{N}}=69$ mOhm at
1.7 K (a) and $R_{\rm{sg}}=71$ mOhm and $R_{\rm{N}}=64$ mOhm at
4.3 K (b).}
\end{center}
\end{figure}

\begin{figure}[ht]
\begin{center}
\includegraphics[width=300pt]{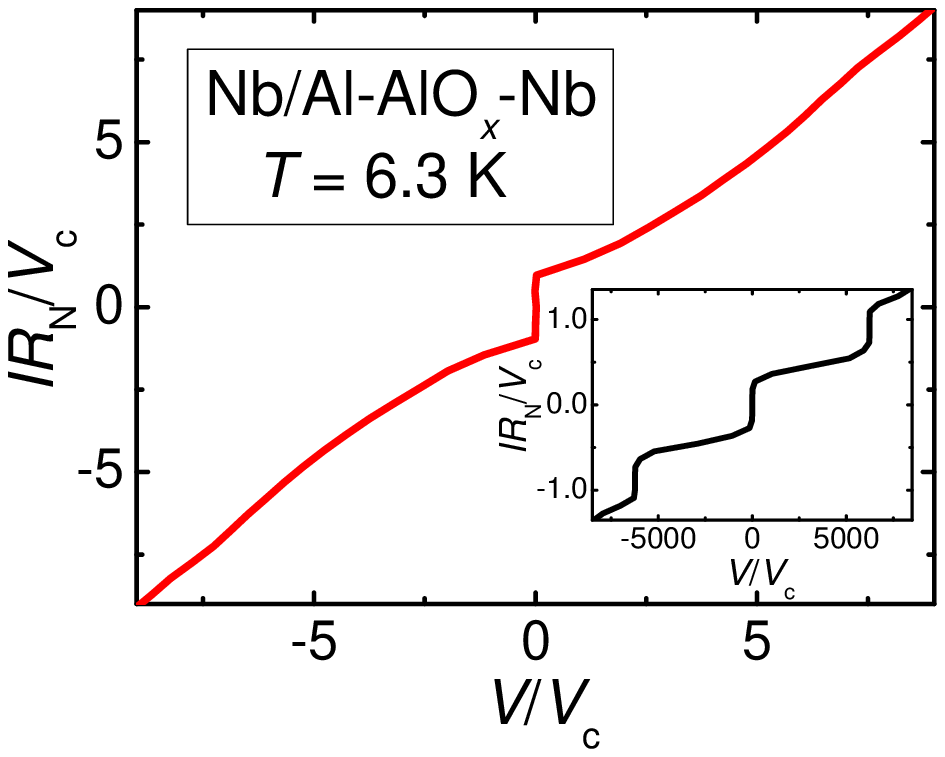}
\caption{ Typical current-voltage characteristic of a
Nb/Al-AlO$_x$-Nb junction with 110 nm-thick Al interlayer (the
main panel) and that of a binary-divided array of the same 8192
SNIS junctions under 73 GHz microwave irradiation (the inset).
Critical currents $I_{\rm {c}}$ and the products $V_{\rm
{c}}=I_{\rm {c}}R_{\rm {N}}$ were about 2.2 mA and 0.2 mV,
respectively at the operating temperature 6.3 K. \label{fig2} }
\end{center}
\end{figure}

Electrical measurements have been performed with a conventional
four-terminal dc technique below critical temperatures of Nb/Al
bilayers which were changed from 8 to 9 K for different $d_{\rm
{Al}}$. All samples have exhibited supercurrents with values of 6
- 15 mA at 1.7 K, 3 - 8 mA at 4.3 K, and 1 - 5 mA at 6.0 K (left
insets in figure~1). A typical current-voltage single-valued
characteristic of a Nb/Al-AlO$_x$-Nb junction at finite
temperatures is shown in the main panel of figure~2. It is well
known~\cite{Likharev} that the subgap current $I_{\rm {sg}}$ is a
sum of three terms: a quasiparticle part $I_{\rm {qp}}$, a
superconducting pair current $I_{\rm {c}}$, and an interference
contribution. The latter two contributions can be made arbitrarily
small if the critical current $I_{\rm{c}}$ is reduced by applying
a magnetic field $B$. Then $I_{\rm{sg}} \approx I_{{\rm{qp}}}$. In
our measurements of dissipative current-voltage characteristics,
magnetic fields $B$ up to 50 mT have been applied to the
Nb/Al-AlO$_x$-Nb samples through a suitable coil. The
$I_{\rm{qp}}-V$ curves have not been changed for two
configurations of the magnetic fields (parallel and orthogonal to
the current flow). The normal resistance $R_{\rm{N}}$ was
determined from a linear fit to current-voltage curves with and
without supercurrents at voltages above 1 mV (see figure~1) and
the results of both estimations were in a good agreement with each
other. It also agrees with the data for Nb-based
trilayers~\cite{Blamire} where specific conductance of thermally
oxidized AlO$_x$ tunnel barriers was above
10$^7$~(Ohm$\cdot$cm$^2$)$^{-1}$ for lowest oxygen exposure doses
similar to ours. Subgap Ohmic resistance $R_{\rm{sg}}$ was
extracted from experimental data as a slope of a best-fit linear
regression line for quasiparticle curves in the interval from 0 to
0.2 mV where the subgap current increases near linearly with $V$
(right insets in figure~1). The ratios $R_{\rm{sg}}/R_{\rm{N}}$ at
4.2 K were always of the order of unity as was found recently in
our work on Nb/Al-AlO$_x$-Nb junctions with ultra-thin
barriers~\cite{LacquanitiJAP} and before it for high-transparency
symmetric Nb-AlO$_x$-Nb samples in~\cite{Blamire}. Complementing
our previous 4.2 K data for $R_{\rm{sg}}$ in SNIS
junctions~\cite{LacquanitiJAP}, in this work we have measured
dissipative $I_{\rm{qp}}-V$ curves also at 1.7 and 6.0 K and
increased the upper value of $d_{\rm{Al}}$ to 150 nm.

\begin{table*}[!]
\begin{center}
\caption{\label{tab:1}Measured parameters for Nb/Al-AlO$_x$-Nb
junctions with different Al thicknesses.}
\begin{tabular}{|c|c|c|c|c|c|c|c|}
\hline
$d_{\rm{Al}}$ & $I_{\rm{c}}(B=0),$ & $R_{\rm N}$, & $R_{\rm N}$, & $R_{\rm N}$, & $R_{\rm {sg}}/R_{\rm N}$. & $R_{\rm {sg}}/R_{\rm N}$. & $R_{\rm {sg}}/R_{\rm N}$.\\
(nm) & 1.7 K & 1.7 K & 4.3 K & 6.0 K & 1.7 K & 4.3 K & 6.0 K\\
& (mA) & (mOhm) & (mOhm) & (mOhm) & & & \\
\hline
45 & 7.01 & 136 & 136 & 132 & 1.39 & 1.36 & 1.14 \\
\hline
57 & 6.13 & 177 & 177 & 175 & 1.31 & 1.29 & 1.07 \\
\hline
110 & 11.92 & 69 & 64 & 63 & 1.38 & 1.11 & 1.05 \\
\hline
115 & 14.37 & 82 & 80 & 80 & 1.30 & 1.15 & 1.03 \\
\hline
142 & 6.87 & 181 & 178 & 177 & 1.27 & 1.13 & 1.03 \\
\hline
\end{tabular}
\end{center}
\vspace{-0.6cm}
\end{table*}

Below we analyze experimental data for five representative SNIS
junctions with very different values of $d_{\rm{Al}}$,
$R_{\rm{N}}$, $I_{\rm{c}}$, and products $I_{\rm{c}}R_{\rm{N}}$.
It means that the samples had dissimilar Nb/Al interface
resistances, as well as oxide barrier parameters, and our aim was
to compare measured ratios $R_{\rm{sg}}/R_{\rm{N}}$ with those
calculated using the universal distribution (1). Table 1
summarizes experimental characteristics of the samples whereas the
main panel of figure~1 shows measured $I-V$ characteristics for
one of the SNIS devices (dots) with two linear fits corresponding
to $I=V/R_{\rm{N}}$ and $I=V/R_{\rm{sg}}$.

\section {Discussion}

Conventional theory of tunneling effects in superconducting
junctions~\cite{Wolf} leads to the following conclusions relating
the $I-V$ curves: first, at 1.7 K for $V\simeq 0.2$ mV the ratio
$R_{\rm{sg}}/R_{\rm{N}}$ should be of the order of 10$^5$-10$^6$
in Nb-I-Nb samples and 10$^2$-10$^3$ in Nb/Al-I-Nb junctions, and
second, the ratio has to decrease (exponentially for SIS contacts)
with increasing temperature up to $\geq10^2$ and $\leq10^2$ at 4.3
K in Nb-I-Nb and Nb/Al-I-Nb devices, respectively. Our samples do
not follow this behavior and exhibit (i) the ratio
$R_{\rm{sg}}/R_{\rm{N}}$ of the order of unity at all temperatures
and (ii) a very weak temperature effect.

To conclude that these observations cannot be attributed to
pinholes, we should prove a sinusoidal current-phase relation
$I_{\rm s}(\phi )$ for our junctions. One of the ways to check it
is to detect Shapiro steps in the $I-V$ curves under microwave
irradiation. Our experiments on SNIS single junctions rf radiated
with frequency 73.5 GHz have displayed a clear harmonic step
structure in the current - voltage characteristics up to 8.3 K
without any subharmonic features~\cite{Lacquaniti2011}. In the
inset of figure~2 we show Shapiro steps at $T = 6.3$ K for binary
divided series arrays of 8192 Josephson SNIS junctions with
$d_{\rm{Al}}=110$ nm, a part of the samples used for subgap
resistance measurements.

It should be noticed that similar results for the ratio $R_{\rm
{sg}}/R_{\rm {N}}\simeq 1$ and its universality have been already
observed in symmetric Nb-AlO$_x$-Nb junctions and interpreted in
the paper~\cite{Naveh} following calculations~\cite{Averin,Bardas}
where the validity of the Schep-Bauer distribution (1) was assumed
and an effect of multiple Andreev reflections important for high
transparencies $D$ was taken into account. In particular, it
follows from figure~2 in the paper~\cite{Bardas} that in SIS
trilayers the $R_{\rm {sg}}/R_{\rm {N}}$ value averaged over the
voltage interval $\left| V \right| < \Delta _{\rm{Nb}}/e$ should
be near 0.7 (the dashed line in figure~3). Experimental data for
Nb-AlO$_x$-Nb samples (see figures~1 and 2 in the
paper~\cite{Naveh}) did well agree with this estimate. Our aim was
three-fold: (i) to introduce an additional Al interlayer in order
to modify the ratio $R_{\rm {sg}}/R_{\rm {N}}$ and to compare
measured values with a calculated one; (ii) to change the
thickness of the interlayer and to reveal in the experiment
whether the distribution (1) comes from the disordered oxide layer
or it is generated by diffusive charge transport in the N region,
and (iii) to investigate for the first time the temperature
dependence of the ratio $R_{\rm {sg}}/R_{\rm {N}}$.

In figure 3 we show experimental data for five representative
samples which agree well with each other after normalization to
$R_{\rm {N}}$. In contrast to the observations for Nb-AlO$_x$-Nb
devices~\cite{Naveh}, the subgap resistance in our samples was
higher than the normal-state value. To explain the findings, in
figure~3 we show linear fits to results of such numerical
calculations for two limiting cases of NIS and SIS junctions
(dotted and dashed lines, respectively). The $I_{\rm{qp}}-V$ curve
for an NIS trilayer was computed by us and that for an SIS
structure was taken from the paper~\cite{Bardas}. For comparison
we plot the current-voltage characteristic in a normal state as
well (dashed-dotted line).

We have also performed numerical simulations of the subgap
resistance in SNIS junctions with arbitrary transparency $D$ of
the interspace between two superconducting electrodes and averaged
the $I_{\rm{qp}}-V$ curves with the relation (1). Since our
approach to the problem is based on numerical methods developed
earlier ~\cite{Hurd1996,Hurd1997,Averin,Bardas,Bratus,Hurd1999},
we shall only briefly outline the main points of the calculations.
The Nb/Al-AlO$_x$-Nb system studied is considered as an asymmetric
S$_1$IS$_2$ junction where S$_1$ is the S/N bilayer. Proximized Nb
layer induces in a thin Al film a superconducting order parameter
which in the general case is a function of $\varepsilon $, the
energy of a quasiparticle excitation in a superconductor measured
from its Fermi level. The proximity effect changes the probability
amplitude of a process of Andreev scattering $a_1(\varepsilon )$
from Al/Nb bilayer comparing with that from a bulk superconductor
and this is just our main modification of the calculation scheme
proposed in the papers referred above. The probability amplitude
$a_1(\varepsilon )$ can be found if the ratio of modified and
normal Green's functions $F$ and $G$ of a superconductor $\Phi
(\omega )=\omega F(\omega )/G(\omega )$ ($\omega $ is the
Matsubara frequency) is known. Note that in a bulk superconductor
$\Phi (\omega )$ is equal to a constant order parameter $\Delta
$~\cite{Golubov2004}. The function $\Phi (\omega )$ of a dirty
normal metal placed in proximity with a superconductor can be
calculated using Usadel equations with corresponding boundary
conditions~\cite{Golubov2004}. In the following, we use a simplest
approximation for the function $\Phi _{\rm {Al}}(\omega )$ in the
Nb/Al bilayer $\Phi _{\rm{Al}} (\omega ) = \Delta _{\rm{Nb}}/(1 +
\gamma \sqrt {\omega ^2  + \Delta _{\rm{Nb}} ^2 } /\Delta
_{\rm{Nb}})$ with a fitting parameter $\gamma $ and the energy gap
of Nb $\Delta_{\rm{Nb}}$~\cite{Golubov1995}. The value of $\gamma
$ can be found by equating the gap magnitude in the Al interlayer
calculated numerically to that found by us experimentally (for
Al-film thicknesses studied its value at 1.7 K can be roughly
estimated as 0.4 meV~\cite{LacquanitiJPCS}). Then we are dealing
with an S$_1$IS$_2$ junction where $a_1 (\omega ) = i\left(
{\omega - \sqrt {\omega ^2 + \Phi _{{\rm{Al}}} ^2 (\omega )} }
\right)/\Phi _{{\rm{Al}}} (\omega )$ and $a_2 (\omega ) = i\left(
{\omega  - \sqrt {\omega ^2  + \Delta _{{\rm{Nb}}} ^2 } }
\right)/\Delta _{{\rm{Nb}}}$. At first, we consider a single-mode
channel with a fixed transparency $D$ between S$_1$ and S$_2$
electrodes. Its length is assumed to be much smaller than elastic
and inelastic lengths. If so, we can describe the transport across
the device in terms of Andreev-reflection amplitudes $a_{1,2}
(\omega )$ and scattering characteristics of the barrier, namely,
its transmission $t$ ($\left| t \right|^2 = D$) and reflection $r$
($\left| r \right|^2  = 1 - D$) probability amplitudes. The energy
of an electron-like quasiparticle going across the barrier is
increased by $eV$ each time when it transfers the classically
forbidden region, for example, from left to right, whereas that of
an Andreev-reflected hole-like excitation increases moving in the
opposite direction to the original electron. These scattering
events will continue back and forth in the interspace between
S$_1$ and S$_2$ superconductors and it is just a process of
multiple Andreev reflections when each round trip of an
electron-like quasiparticle increases its energy with a value of
$2eV$. As a result, from each side of the barrier we get an
infinite set of scattering states with different energies shifted
by $2eV$~\cite{Averin,Bardas}. By relating the wave functions at
the two sides of the scattering region via scattering $t$ and $r$
amplitudes, we obtain recurrence relations for amplitudes of
electron-like and hole-like wave functions in an asymmetric
S$_1$IS$_2$ junction~\cite{Hurd1996,Hurd1999} which have been
solved numerically. After obtaining the scattering states, we
calculate the electrical current $I_{\rm{qp}}(D,V)$ as a function
of the applied voltage and the parameter $D$, taking into account
that the contribution to the current from injected hole-like
quasiparticles is equal to that from electrons. The total
current-vs-voltage characteristic can be represented as a sum of
independent contributions from individual transverse modes with a
known distribution of their transmission probabilities $\rho (D)$:
$I_{\rm{qp}}(V)=\int\limits_0^1 d D\rho (D)I_{\rm{qp}}(D,V)$.
\begin{figure}[ht]
\begin{center}
\includegraphics[width=320pt]{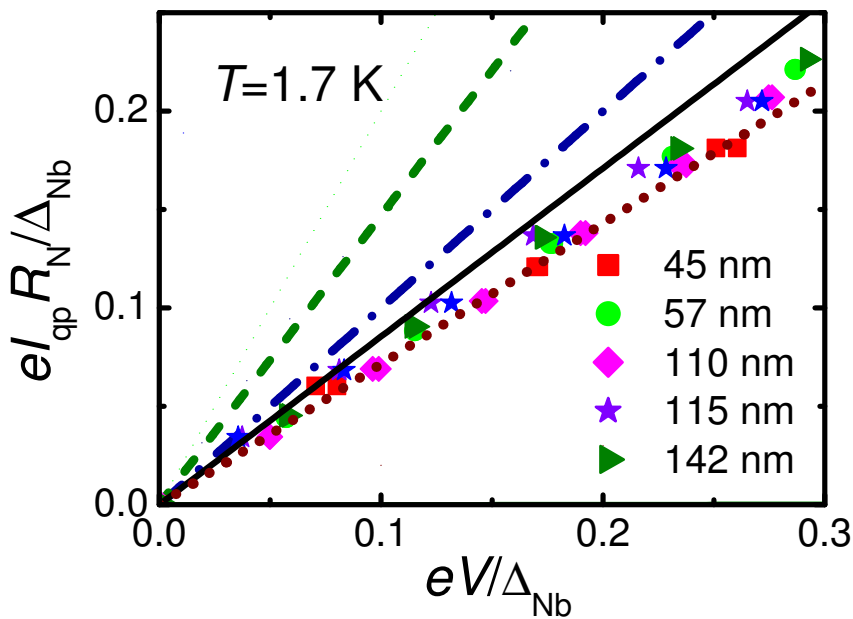}
\caption{(Color online) Quasiparticle current-voltage
characteristics of five Nb/Al-AlO$_x$-Nb junctions with different
thicknesses of the Al interlayer (indicated in the figure) at 1.7
K (symbols) compared with linear fits to calculated dependences
for NIS, SIS~\cite{Bardas}, SNIS, and NIN junctions with a
disordered barrier layer (dotted, dashed, solid, and dashed-dotted
lines, respectively).} \label{fig3}
\end{center}
\end{figure}

Reasonable agreement between the calculated $R_{\rm {sg}}/R_{\rm
N}$ ratio and experimental data for Nb/Al-AlO$_x$-Nb junctions in
figure~3 proves that, independently on the presence and on the
thickness of the Al interlayer, the main spread of the
transmission coefficients is related to the oxide barrier.
Moreover, the Schep-Bauer distribution (1) not only qualitatively
but also quantitatively describes the experimental data. As was
expected, $R_{\rm {sg}}/R_{\rm N}$ values for the Nb/Al-AlO$_x$-Nb
system are closer to those for NIS devices than to those for
symmetric SIS junctions. From figure~3 it is clear that the
approximation for $\Phi _{\rm {Al}}(\omega )$ used by us slightly
exaggerates the proximity effect in Al.

Let us stress that while the $R_{\rm {sg}}/R_{\rm N}$ value is
almost the same for different $d_{\rm {Al}}$ at 1.7 K, its
temperature behavior, as follows from table 1, is strongly
different. For the lowest $d_{\rm {Al}}$, the deviation of the
ratio value at 4.3 K from that at 1.7 K is tiny whereas for larger
thicknesses the $R_{\rm {sg}}/R_{\rm N}$ value tends to unity. It
can be explained as a result of a very different temperature
behavior of the energy gap induced in an Al interlayer. It follows
from figure~3 in the paper~\cite{LacquanitiJPCS} that the
temperature suppression of the induced gap and, hence, the
asymmetry of SNIS superconducting junctions become stronger for
thicker interlayers. As a result, in the subgap voltage range a
strong peak evolves in $I-V$ curves for comparatively small
barrier transparencies (see figure~5 in the paper~\cite{Hurd1997})
and this thermally induced decrease of the subgap resistance is
more effective for larger $d_{\rm {Al}}$.

Disappearance of a hysteretic $I-V$ response in Nb/Al-AlO$_x$-Nb
Josephson junctions when rising temperature from 1.7 K to 4.3 K
(compare left insets in figures~1a and 1b) is partly an effect of
the temperature influence on the $R_{\rm {sg}}/R_{\rm N}$ ratio.
It is well known~\cite{Likharev} that the Josephson-device
operation regime is governed by the McCumber-Stewart damping
parameter
\begin{equation}
\beta _{\rm c} = \frac{{2e^2 }}{\hbar }\left( {I_{\rm{c}}
R_{\rm{N}}^2C}\right)\left({R_{\rm{sg}}/R_{\rm{N}} } \right)^2
\end{equation}
with $C$, the junction capacitance. If $\beta _{\rm c}$ is less
than unity the current-voltage characteristic of the junction in a
superconducting state is single valued like that shown in
figure~1b. Otherwise, the response is hysteretic with switching
$I_{\rm c}$ and retrapping $I_{\rm r}$ currents (figure~1a). We
can estimate the $\beta _{\rm c}$ value for a representative
junction shown in figure~1 using the Zappe's formula~\cite{Zappe}
$\beta _{\rm c}=[2 -(\pi -
2)(I_{\rm{r}}/I_{\rm{c}})]/(I_{\rm{r}}/I_{\rm{c}})^2$. At 1.7 K we
get $\beta _{\rm c}=2.21$. After increasing temperature to 4.3 K,
the critical current decreased from 11.0 to 6.9 mA, As a result,
the McCumber-Stewart parameter due to Eq.~(2) should reduce to
1.38. With an additional factor, temperature-induced decline of
the $R_{\rm {sg}}/R_{\rm N}$ ratio squared (Table 1), we obtain a
final value $\beta _{\rm c}=1.07$ in agreement with the experiment
showing a very small hysteresis in the current-voltage
characteristics at 4.3 K (the left inset in figure~1b).

Let us now discuss the origin of universality of the distribution
(1). We start with a simple parameterization $D(Z) =
(1 + Z^2)^{-1}$ which follows directly from the delta-functional
approximation for a potential barrier in a one-dimensional NIN
trilayer~\cite{Blonder} where $Z=k_{\rm{F}}\int _0^d V_{\rm{B}}(x) dx
/E_{\rm{F}}$, $V_{\rm{B}}(x)$ and $d$ are the local barrier height
and the thickness of the insulating layer, $k_{\rm{F}}$ and
$E_{\rm{F}}$ are the Fermi wave vector and the Fermi energy in a
metallic electrode. To find the probability function for a transformed variable,
we use a standard relation $\rho (Z)=\rho (D)(dD/dZ)$ and obtain that
\begin{equation}
\rho (Z)=2\hbar G_{\rm{N}}/e^2=\rm {const}
\end{equation}
with Z uniformly ranging from zero to infinity. Thus, for very
thin barriers Eq. (1) is equivalent to the assumption that an
integral of the potential barrier height taken along an electron
path inside the classically forbidden region is a uniform random
variable (note that in a strongly disordered insulating layer such
a path can significantly exceed the nominal barrier thickness $d$
due to elastic subbarrier defect scatterings).

Even more, we cam conclude that the Schep-Bauer formula (1) for the transparency $D$ is not limited
to this assumption but is valid, independently on the physical nature, when (i) the transparency may
be represented as a one-parameter Lorentzian and (ii) this parameter is uniformly distributed from very small up
to very large values (see the related discussion concerning
spatial distribution of barrier defects in the paper by Il'ichev
{\it {et al.}}~\cite{Il'ichev}). It is interesting that before the
work~\cite{Schep} the same relation (1) was derived for a
quasi-ballistic double-barrier INI interspace with two identical
uniform insulating layers~\cite{Melsen}, a system which is
physically very different from a thin disordered dielectric film.
In our opinion, the reason for coincidence between the two systems
consists in the fact that, for a finite thickness of the N
interlayer $d_{\rm N}$ the transmission coefficient $D$ is also of
a Lorentzian-like form. To show it, we express the transmission
amplitude $\widetilde{t}$ across the I$_1$NI$_2$ transition region
as a sum of a geometric series
\begin{equation}
\tilde t = t_1 \exp (i\varphi )t_2 (1 + r_1 \exp (2i\varphi )r_2 +
...) = \frac{{t_1 \exp (i\varphi )t_2 }}{{1 - r_1 \exp (2i\varphi
)r_2 }},
\end{equation}
where $\varphi (\theta) =kd_{\rm N} \rm {cos(\theta)}$ is the
phase shift acquired by an electron with a wave number $\bf {k}$
traveling between two interlayer boundaries; $t_{1,2}(\theta)$ and
$r_{1,2}(\theta)$ are transmission and reflection amplitudes for
I$_1$ and I$_2$ insulating layers, respectively; $\theta $ is the
injection angle between the vector $\bf {k}$ and the normal to
layer interfaces. Then the probability of charge transition across
the double-barrier I$_1$NI$_2$ trilayer can be expressed as
$D(\theta )=(1 + \tilde Z^2(\theta ))^{ - 1}$ with
\begin{equation}
\tilde Z = \sqrt {{2 - \left| {t_1 } \right|^2  - \left| {t_2 }
\right|^2  - 2{\mathop{\rm Re}\nolimits} \left\{ {r_1 r_2 \exp
(2i\varphi )} \right\}} }/ \left( {\left| {t_1 } \right|\left|
{t_2 } \right|} \right).
\end{equation}
For a sufficiently large
thickness $d_{\rm N}$ and symmetrical INI structure with
low-transmission barriers $\tilde Z(\theta )$ is a rapidly
oscillating function, which changes periodically from zero in the
resonance state to very high values. This very wide spread of the
parameter $\tilde Z(\theta )$ is the reason why the two
distributions coincide.

Now we transfer to experimental facts supporting the statement
about a broad spread of local barrier transparencies in very thin
dielectric layers and explaining its origin. First, we refer to a
recent paper by Welander \textit{et al}~\cite{Welander}, where two
ways of Al-oxide barrier formation in Nb/Al-AlO$_x$-Nb junctions
were employed. The first process based on the conventional
diffusion-limited oxidation of the Al layer yielded Josephson
tunnel junctions with significant subgap leakage whereas the
subgap currents in devices with layer-by-layer grown barriers
agreed well with a standard tunneling theory. According to the
paper~\cite{Welander} the extra conductance, most probably, comes
from defects in the diffused Al oxide caused by room-temperature
thermal oxidation of Al. These defects are known to be positively
charged oxygen vacancies~\cite{Tan}. Another work~\cite {Julin}
supporting the idea about the decisive role of oxygen vacancies is
dealing with low-frequency noise measurements in Al-AlO$_x$-Al
tunnel junctions. It was found that vacuum thermal annealing
strongly reduces the $1/f$ noise level in the Al-based devices
(sometimes, up to an order of magnitude). Since the $1/f$ noise
phenomenon in metal-insulator-metal tunnel junctions is definitely
related to slow filling and emptying of localized electron states
in the barrier~\cite{Rogers}, the finding by Julin \textit{et
al.}~\cite{Julin} can be understood as a reduction of the charge
traps number within the AlO$_x$ layer by the annealing procedure.

Using conductive atomic force microscopy technique, Kim \emph{et
al.}~\cite{Kim} studied local tunneling properties of ultra-thin
MgO films. Topographic maps showed that MgO layers were very flat
whereas local tunnel current maps were strongly inhomogeneous with
a number of current hotspots without any correlation between the
two maps for the same sample. The most probable explanation can be
that the tunnel current inhomogeneity arises from fluctuations in
barrier height but not structural fluctuations. Comparing the maps
before and after adding O$_2$ to the Ar plasma during MgO growth,
the authors of the paper~\cite{Kim} argued that the inhomogeneity
originates from oxygen vacancies. Elimination of certain oxygen
defect populations leads to improved barrier heights with reduced
spatial variations. Additional argument relating the presence of
defect states in a very thin MgO tunnel barrier comes from the
non-linearity of $I-V$ characteristics of corresponding magnetic
tunnel junctions. The presence of oxygen vacancies in MgO leads to
a new tunneling transport mechanism within the film when a charge
is transported across the classically forbidden region via
boson-induced tunneling jumps between randomly distributed
localized states~\cite {Glazman}. Combination of direct and
inelastic tunneling events through the defects should lead to
anomalous voltage dependence of the differential conductance in
the form of a sum of power functions of $V$ with different
non-integer exponents. Such a behavior was just observed in the
paper~\cite{Gokse} for magnetic tunnel devices with MgO barriers.
Very recent examples proving relation of anomalous properties of
thin oxide layers to oxygen-originated inhomogeneities are given
in the papers~\cite{Perevalov,Arhammar}. Together with those
discussed above they show that defects in disordered alumina are
definitely related to oxygen although their detailed structure
remains unclear (of course, this statement is material-dependent).

But which physical quantity does mainly fluctuate in the
disordered nanometer-thick oxide layers - the thickness $d$ as was
assumed in~\cite{Zorin} or the average barrier height? By other
words, in which way do the oxygen defects influence the tunnel
current? The authors~\cite{Choi} paid attention to the role of
metal-induced evanescent electronic states in the nearest to metal
electrodes two-three layers of insulator unit cells. Inevitably
random fluctuations in the electronic potential due to formation
of oxygen defects should localize a substantial fraction of the
evanescent states and, in our opinion, it is just the origin of
strong local barrier strength variations within the insulating
barrier. Let us emphasize that this scenario is exceptionally
important just for ultra-thin oxide barriers as ours, since their
thickness is expected to be of several unit cells and they
directly contact metallic films from both sides.

In conclusion, highly-conductive asymmetric Nb/Al-AlO$_x$-Nb
junctions show strong subgap leakage currents as was observed
earlier in symmetric Nb-AlO$_x$-Nb trilayers~\cite{Naveh} with
high-transparency potential barriers similar to ours. Moreover, we
have found that at 1.7 K the ratio of subgap to normal resistances
does not greatly change for samples with Al-interlayer thicknesses
ranging from 40 to 140 nm. This finding was explained as an
indication of a universal distribution of barrier transparencies
in ultra-thin Al oxide films. A noticeably reduced subgap leakage
comparing with corresponding Nb-AlO$_x$-Nb trilayers~\cite{Naveh}
is in agreement with analogous finding for Nb/Al-AlO$_x$-Al
samples with a 5 nm-thick interfacial Al film~\cite{Im}. We do
agree with the authors~\cite{Im} that the magnitude of the leakage
does not significantly depend on a particular material of the
metallic layers. The temperature impact on the subgap junction
resistance can be understood taking into account different
temperature behavior of a superconducting energy gap induced in
the proximized Al interlayer. Our results clearly indicate that in
the case of ultra-thin and disordered Al-oxide films the main
effect comes from the distribution of local oxide transparencies.
Suppression of the subgap current with introducing the Al
interlayer is caused by reducing the energy gap of one of the
electrodes whereas the key factor, the universal distribution (1),
remains almost the same as in Nb-AlO$_x$-Nb structures. We have
argued that its presence originates from a very broad and
homogeneous distribution of local barrier heights generated by
oxygen vacancies within the Al-oxide film. If the defect
distribution is more or less uniform~\cite{Il'ichev}, the integral
of the potential barrier height along an electron path inside the
barrier would be a uniform random variable. Of course, this
phenomenon is not scale invariant. Since the depth of the
penetration of metal-induced evanescent electronic states is
limited to two-three atomic layers~\cite{Choi}, leakage currents
should be more pronounced for extremely thin insulating layers.
The way to suppress it is to use dielectrics with lower band
forbidden gap~\cite{Endo}. Because the effect is self-averaging,
it will be more pronounced in devices with comparatively large
junction area like ours. The latter remark explains why a
noticeable reduction of the subgap current can be achieved by
splitting a high-transparency micrometer-scale SIN heterostructure
into several submicron sub-junctions~\cite{Zorin}.

\section*{Acknowledgements}

This work was supported by a grant from Regione Piemonte. One of
us (M.B.) would like to thank   the National Institute of
Metrological Research in Turin, Italy for support during the
course of the work.

\end{document}